\documentclass[final,3p,times]{elsarticle}

\usepackage{amssymb}
\usepackage{amsthm}
\usepackage{amsmath}
\usepackage{multirow}
\usepackage{listings}
\usepackage{color}
\newcounter{bla}

\journal{Computer Physics Communications}

\begin{document}

\begin{frontmatter}

\title{PYATB: An Efficient Python Package for Electronic Structure Calculations Using Ab Initio Tight-Binding Model}

\author[a,c]{Gan Jin}
\author[a,c]{Hongsheng Pang}
\author[a,c]{Yuyang Ji}
\author[a,c]{Zujian Dai}
\author[a,b,c]{Lixin He \corref{author}}

\cortext[author] {Corresponding author.\\\textit{E-mail address:} helx@ustc.edu.cn}
\address[a]{CAS Key Laboratory of Quantum Information, University of Science and Technology of China, Hefei, Anhui, 230026, China}
\address[b]{Institute of Artificial Intelligence, Hefei Comprehensive National Science Center,
		Hefei, Anhui, 230026, People's Republic of China}
\address[c]{Hefei National Laboratory, University of Science and
		Technology of China, Hefei, Anhui,  230088, People's Republic of China}

\begin{abstract}

We present PYATB, a Python package designed for computing band structures and related properties of materials using the ab initio tight-binding Hamiltonian. The Hamiltonian is directly obtained after conducting self-consistent calculations with first-principles packages using numerical atomic orbital (NAO) bases, such as ABACUS. The package comprises three modules: Bands, Geometric, and Optical. In the Bands module, one can calculate essential properties of band structures, including the partial density of states (PDOS), fat bands, Fermi surfaces, and Weyl/Dirac points. The band unfolding method is utilized to obtain the energy band spectra of a supercell by projecting the electronic structure of the supercell onto the Brillouin zone of the primitive cell. With the Geometric module, one can compute the Berry phase and Berry curvature-related quantities, such as electric polarization, Wilson loops, Chern numbers, and anomalous Hall conductivities. The Optical module offers a range of optical property calculations, including optical conductivity and nonlinear optical responses, such as shift current and Berry curvature dipole.

\end{abstract}


\end{frontmatter}



{\bf PROGRAM SUMMARY}

\begin{small}
\noindent
{\em Program Title: PYATB}                                          \\
{\em CPC Library link to program files:} (to be added by Technical Editor) \\
{\em Developer's repository link: https://github.com/pyatb/pyatb} \\
{\em Code Ocean capsule:} (to be added by Technical Editor)\\
{\em Licensing provisions: GPLv3} \\
{\em Programming language: C++, Python}                                   \\
{\em Nature of problem: This program is to study the electronic structure, electronic polarization, band topological properties, topological classification,  linear and nonlinear optical response of solid crystal systems.} \\
{\em Solution method: Based on the tight binding method to solve the band structure, the Wilson loop is used to classify the topological phases, and the optical response is calculated by Berry curvature and Berry connection.} \\

\end{small}

\section{Introduction}
\label{introduction}

Electronic band structures are critical in determining the physical properties of solids, including their optical, transport, and topological properties. For instance,  in materials with nontrivial topological properties, such as topological insulators~\cite{hsieh_topological_2008, zhang_topological_2009, xia_observation_2009}, topological crystalline insulators~\cite{fu_topological_2011}, Dirac~\cite{wang_three-dimensional_2013, wang_dirac_2012} and Weyl semimetals~\cite{wan_topological_2011, weng_weyl_2015, soluyanov_type-ii_2015}, and nodal-line semimetals~\cite{burkov_topological_2011, yu_topological_2015, bian_drumhead_2016}, calculations of the Berry curvature~\cite{chang_berry_2008}, Chern number~\cite{thouless_quantized_1982}, and Wannier charge centers~\cite{soluyanov_computing_2011} are crucial for understanding these topological states.
Moreover, recent studies have revealed that the Berry connection and Berry curvature also play essential roles in various nonlinear optical effects~\cite{morimoto_topological_2016, nagaosa_concept_2017}.
Kohn-Sham density functional theory \cite{hohenberg_inhomogeneous_1964, kohn_self-consistent_1965}  is a vital tool for calculating band structures and their associated properties. However, these calculations typically require a large number of $\mathbf{k}$ points, making them computationally demanding. The tight-binding model offers an efficient approach to investigating electronic structures in materials.
In plane wave-based codes, the Maximally Localized Wannier Functions (MLWF) method~\cite{Marzari_maximally_2012} is used to construct the ab initio tight-binding model, where the parameters are determined from first-principles calculations. The MLWF method has been implemented in the Wannier90~\cite{mostofi_wannier90_2008} code, which provides interfaces with many widely used first-principles software, such as Quantum-Espresso \cite{giannozzi_quantum_2009}, VASP \cite{kresse_efficient_1996, kresse_efficiency_1996}, and ABINIT \cite{gonze_the-abinit_2020} etc.
Several packages, including Wannier90~\cite{mostofi_wannier90_2008}, Z2pack~\cite{gresch_z2pack_2017}, and WannierTools~\cite{wu_wanniertools_2018} use the MLWF based ab initio Hamiltonian as the postprocess to calculate various properties of materials, such as the band structures, band geometric and topological properties, and optical properties.
However, obtaining high-quality WFs for large systems can be computationally demanding, and in some cases, the symmetry of the WFs may not be preserved, \cite{mostofi_wannier90_2008} which could potentially lead to incorrect results for
properties that are sensitive to the symmetry of the system.

However, if the numerical atomic orbitals (NAO) are used as the basis set in the first-principles package, the ab initio tight-binding Hamiltonian can be naturally generated after the self-consistent calculations, avoiding the process to construct MLWF and can preserve the correct symmetries of the systems.
In this paper, we present the PYATB package, which is designed for computing band structures and related properties of materials using the ab initio tight-binding Hamiltonian on the NAO bases. The package includes three modules that enable the computation of different material properties. The {\tt Bands module} calculates basic properties of band structures, including the fat bands, PDOS and Fermi surface etc. Moreover, it can also detect the Dirac/Weyl points and nodal lines in topological semi-metals. Additionally, the module includes a band-unfolding method that enables the calculation of energy band spectra for large supercells.
The {\tt Geometric module} focuses on geometric and topological properties of the materials, while the {\tt Optical module} computes linear and nonlinear optical properties.

PYATB provides a user-friendly interface for carrying out calculations on specific functionalities through an {\tt Input} file. It can also be used as a Python module for customized function calculations by leveraging APIs, such as Berry curvature, Berry connection, and velocity matrix, which enables deeper exploration and reduces the burden of software development.
Currently, PYATB has an interface with the first-principles package ABACUS~\cite{li_large-scale_2016}, but it is straightforward to construct the interface with other NAO-based first-principles softwares.

The rest of the paper is organized as follows. In Sec.~\ref{sec:methhod}, we introduce the capabilities, theory, and implementation of the PYATB package. Section~\ref{sec:install} covers the installation process of PYATB and how to run it. In Section~\ref{sec:examples}, we provide several examples that demonstrate the capabilities of PYATB. We summarize in Sec.~\ref{sec:summary}.

\section{Capabilities and method}
\label{sec:methhod}

In this sections, we provide a brief overview of the capabilities,
and some basic theories and implementations of these features in PYATB.

\subsection{Capabilities}
\begin{table*}[htbp]
	\centering
	\caption{Main capabilities of PYATB}
	\begin{tabular}{c | c}
		\hline
		Module & Functions \\
		\hline
		\multirow{7}{*}{Bands } & band structure \\
		& band unfolding \\
		& fermi energy and fermi surface \\
		& find nodes \\
		& DOS and PDOS \\
		& fat band \\
		& spin texture \\
		\hline
		\multirow{6}{*}{Geometric} & Wilson loop \\
		& electric polarization \\
		& Berry curvature \\
		& anomalous Hall conductivity \\
		& Chern number \\
		& Chirality \\
		\hline
		\multirow{4}{*}{Optical } & JDOS \\
		& optical conductivity and dielectric function \\
		& shift current \\
		& Berry curvature dipole \\
		\hline
	\end{tabular}
	\label{table:capabilities}
\end{table*}
PYATB is a powerful tool for calculating and analyzing the electronic band structure of materials. It provides three major modules: the {\tt Bands module}, the {\tt Geometric module}, and the {\tt Optical module}. The capabilities of these modules are summarized in Table~\ref{table:capabilities}.

The {\tt Bands module} includes seven functions:

\begin{itemize}
\item {\tt Band structure:} Allows users to calculate the energy bands and wave functions using three different $\mathbf{k}$-point modes: k-point, k-line, and k-mesh.

\item {\tt Band unfolding:} Calculates the spectra weight by unfolding the energy bands of a supercell into the Brillouin zone
(BZ) of the primitive cell.

\item {\tt Fermi energy and Fermi surface:} Calculates the Fermi energy at a given temperature and plots the Fermi surface.

\item {\tt Find node:} Allows users to search for degenerate points of the energy bands in the BZ within a specified energy window. This function can be used to find the Weyl/Dirac points in Weyl/Dirac semimetals.

\item {\tt DOS and PDOS:} Calculates the density of states (DOS) and partial density of states (PDOS) of particular orbitals.

\item {\tt Fat band:} Provides the contribution of each atomic orbital to the electronic wave functions at each $\mathbf{k}$-point in the BZ.

\item {\tt Spin texture:} Plots the spin polarization vector as a function of momentum in the BZ.
\end{itemize}

The {\tt Geometric module} calculate the band geometry related properties, which offers six functions, including:

\begin{itemize}
\item {\tt Wilson loop:} Enables users to calculate the $\mathbb{Z}_2$ number by tracking the Wannier centers~\cite{soluyanov_computing_2011} along the Wilson loop.

\item{\tt Electric polarization:} Evaluates the electric polarization in various directions for non-centrosymmetric materials based on the Berry phase theory.

\item{\tt Berry curvature:} Computes the Berry curvature in the BZ.

\item{\tt Anomalous Hall conductivity:} Calculates the anomalous Hall conductivity using Berry curvature.

\item{\tt Chern number:} Calculates the Chern number of a system for any given $\mathbf{k}$-plane.

\item{\tt Chirality:} Examines the chirality of Weyl points by calculating the Berry curvature on a sphere around the $\mathbf{k}$ point.
\end{itemize}

The {\tt Optical module} includes four functions, which are listed below:

\begin{itemize}
\item {\tt JDOS:} Calculates the joint density of states (JDOS), which characterizes both electronic states and optical transitions.

\item {\tt Optical conductivity and dielectric function:} Calculates the frequency-dependent optical conductivity and dielectric function.

\item {\tt Shift current:} Calculates the shift current conductivity tensor for the bulk photovoltaic effect.

\item {\tt Berry curvature dipole:} Calculates the Berry curvature dipole which leads to the nonlinear anomalous Hall effects.
\end{itemize}

\subsection{Ab initio tight binding method on the NAO bases}

PYATB is based on the ab initio tight binding model, where the parameters of the Hamiltonian are generated directly from the self-consistent calculations using first-principles software based on NAO bases, such as ABACUS ~\cite{li_large-scale_2016}. Usually the NAO bases are non-orthogonal.

In a periodic system, the Kohn–Sham equation at a given  $\mathbf{k}$ point can be written as,
\begin{equation}\label{eq:ks}
\hat{H} |\Psi_{n\mathbf{k}} \rangle  = E_{n\mathbf{k}} |\Psi_{n\mathbf{k}}\rangle \, .
\end{equation}
Here $\Psi_{n\mathbf{k}}$ is the Bloch wave function of the $n$-th band, and can be expressed as a linear combination of atomic orbitals,
\begin{equation}\label{eq:eigenv}
|\Psi_{n\mathbf{k}}\rangle = \frac{1}{\sqrt{N}}\sum_{\mu}C_{n\mu}(\mathbf{k}) \sum_{\mathbf{R}}\mathrm{e}^{i\mathbf{k}\cdot\mathbf{R}}|\mathbf{R}\mu\rangle.
\end{equation}
In Eq. (\ref{eq:eigenv}), $|\mathbf{R}\mu\rangle \equiv \phi_{\mu}\left(\mathbf{r}-\tau_{\mu}-\mathbf{R}\right)$ is the $\mu$-th atomic orbital, in the $\mathbf{R}$-th unit cell, and $\tau_{\mu}$ denotes the center position of this orbital.
The composite index $\mu = (\alpha, i, \zeta, l, m)$, where $\alpha$ is the element type, $i$ is the index of the atom of each element type, $\zeta$ is the multiplicity of the radial functions for the angular momentum $l$, and $m$ is the magnetic quantum number.
The coefficient of the NAO is given by $C_{n\mu}(\mathbf{k})$.
In many calculations, the cell periodic part of the Bloch wave functions is used, which is given by,
\begin{equation}
|u_{n\mathbf{k}}\rangle = \frac{1}{\sqrt{N}}\sum_{\mu}C_{n\mu}(\mathbf{k}) \sum_{\mathbf{R}}\mathrm{e}^{i\mathbf{k}\cdot \left(\mathbf{R} - \mathbf{r}\right)}|\mathbf{R}\mu\rangle.
\label{eq:cell_periodic_wf}
\end{equation}

By substituting  Eq. (\ref{eq:eigenv}) into Eq. (\ref{eq:ks}), the Kohn-Sham equation becomes a general eigenvalue problem in the NAO bases,
\begin{equation}
\label{eq:diago_H}
H(\mathbf{k}) C_{n}(\mathbf{k}) = E_{n\mathbf{k}} S(\mathbf{k}) C_{n}(\mathbf{k}),
\end{equation}
where $H(\mathbf{k})$, $S(\mathbf{k})$ and $C_{n}(\mathbf{k})$ are the Hamiltonian matrix, overlap matrix and eigenvectors of the $n$-th band, respectively.

To obtain the $H(\mathbf{k})$ and $S(\mathbf{k})$, we first calculate tight binding Hamiltonian in real space via first-principles softwares based on NAOs , e.g., ABACUS~\cite{li_large-scale_2016},
\begin{eqnarray}
H_{\nu\mu}(\mathbf{R}) &=& \langle \mathbf{0}\nu |  \hat{H} | \mathbf{R}\mu \rangle \label{eq:HR} \, , \\
S_{\nu\mu}(\mathbf{R}) &=& \langle \mathbf{0}\nu | \mathbf{R}\mu \rangle \label{eq:SR}\, .
\end{eqnarray}
Once we have the $H_{\nu\mu}(\mathbf{R}) $ and $S_{\nu\mu}(\mathbf{R})$, we can obtain the Hamiltonian matrix and the overlap matrix at arbitrary $\mathbf{k}$ points using the following relation,
\begin{eqnarray}
H_{\nu\mu}(\mathbf{k}) &=& \sum_{\mathbf{R}} \mathrm{e}^{i\mathbf{k}\cdot\mathbf{R}} H_{\nu\mu}(\mathbf{R}) , \\
S_{\nu\mu}(\mathbf{k}) &=& \sum_{\mathbf{R}} \mathrm{e}^{i\mathbf{k}\cdot\mathbf{R}} S_{\nu\mu}(\mathbf{R}) .
\end{eqnarray}
The band structure and Bloch wave functions can be calculated by solving the general eigenvalue problem in Eq.~(\ref{eq:diago_H}). Note that $H_{\nu\mu}(\mathbf{R}) $ and $S_{\nu\mu}(\mathbf{R})$ can be generated using a relatively  small set of coarse grid $\mathbf{k}$ points. This feature is extremely useful when a large number of $\mathbf{k}$ points are needed to calculate the physical properties. Especially for hybrid functionals, the expensive self-consistent calculations are only required to obtain $H({\bf R})$ at the coarse grid  $\mathbf{k}$ points. Once we have obtained $H({\bf R})$, we can efficiently calculate band structures and associated properties at much denser $\mathbf{k}$ points at the exact same cost as that of LDA and GGA functionals.

In order to calculate the Berry phase and Berry curvature and optical responses, we also need to calculate the dipole matrix between the NAOs,
\begin{equation}
\mathbf{r}_{\nu\mu}(\mathbf{R}) = \langle \mathbf{0}\nu | \mathbf{r} |\mathbf{R}\mu \rangle \label{eq:rR}\,.
\end{equation}
To calculate the dipole matrix, we first expand $\mathbf{r}$ in the spherical coordinate system,
\begin{equation}
\mathbf{r} = r \left(\sin\theta\cos\varphi + \sin\theta\sin\varphi + \cos\varphi\right)\,,
\end{equation}
then express it in terms of the real spherical harmonics,
\begin{equation}
\mathbf{r} = r \times \sqrt{\frac{4\pi}{3}} \left(S^0_1 - S^1_1 - S^{-1}_1\right) \,,
\label{eq:r-decompose}
\end{equation}
where $S^l_m(\theta, \varphi)$ is real spherical harmonics, $l$ and $m$ are the spherical harmonic degree and order, respectively. We calculate $\mathbf{r}_{\nu\mu}(\mathbf{R})$  using Eq. (\ref{eq:r-decompose}) and utilizing the three-orbit double-center integral method~\cite{soler_siesta_2002}.
The dipole matrix $\mathbf{A}^{R}_{\nu\mu}(\mathbf{k})$ can then be obtained by Fourier transform,
\begin{equation}\label{eq:dipole}
\mathbf{A}^{R}_{\nu\mu}(\mathbf{k}) = \sum_{\mathbf{R}}\mathrm{e}^{i\mathbf{k}\cdot\mathbf{R}} \mathbf{r}_{\nu\mu}(\mathbf{R})\,.
\end{equation}

\subsection{Fermi Energy at finite temperatures}

The Fermi distribution function at finite temperature $T$ is given by,
\begin{equation}
f(E,\mu,T)=\frac{1}{1+e^{\left(E-\mu\right)/(k_B T)}}\,.
\end{equation}
The Fermi energy $\mu$ at $T$ can be obtained by solving the following equation,
\begin{equation}
N_{\rm elec}=g\int_{BZ} d\mathbf{k} \sum_n f(E_{n\mathbf{k}},\mu,T) \, ,
\end{equation}
where $N_{\rm elec}$ is the total valence electrons in the unit cell, and $g$ is the spin degeneracy. The integration is over the first BZ.
This integration equation is solved by Newton's method. If the system is an insulator, $\mu$ is given by the valence band maximum (VBM) .

\subsection{PDOS and fat band}

The distribution of electronic states at various energies is characterized by the density of states (DOS), while the partial density of states (PDOS) is a useful tool for analyzing the contribution of individual  atomic orbitals to the DOS. The PDOS of the $\mu$-th orbital can be calculated by projecting the Bloch wave functions onto the atomic orbital,
\begin{equation}
g_{\mu}(E)= \frac{1}{N_{\mathbf{k}}} \sum_{\mathbf{k}} \sum_{n} \langle \Psi_{n\mathbf{k}}  |\phi_{\mu}(\mathbf{k}) \langle \tilde{\phi}_{\mu}(\mathbf{k}) | \Psi_{n\mathbf{k}} \rangle \delta(E - E_{n\mathbf{k}}) \, ,
\end{equation}
where $|\phi_{\mu}(\mathbf{k})\rangle = \frac{1}{\sqrt{N}}\sum_{\mathbf{R}} \mathrm{e}^{i\mathbf{k}\cdot\mathbf{R}}|\mathbf{R}\mu\rangle$, and
$\langle \tilde{\phi}_{\mu}(\mathbf{k})|$=$\sum_{\nu}  S_{\mu\nu}^{-1}(\mathbf{k}) \langle \phi_{\nu}(\mathbf{k}) |$ is the dual function of $|\phi_{\mu}(\mathbf{k})\rangle $. Using Eq. (\ref{eq:eigenv}), the PDOS is calculated as,
\begin{equation}
g_{\mu}(E)= \frac{1}{N_{\mathbf{k}}} \sum_{\mathbf{k}} \sum_{n} \sum_{\nu} C_{n\nu}^{*}(\mathbf{k}) S_{\nu\mu}(\mathbf{k}) C_{n\mu}(\mathbf{k}) \delta(E - E_{n\mathbf{k}}).
\end{equation}

A fat band can provide information about the contributions of specific atomic orbitals or groups of orbitals to the electronic bands of a material at  given $\mathbf{k}$ points. The orbital weight is calculated by projecting the Bloch wave function onto the selected atomic or group of atomic orbitals, which can be calculated in a similar way to that of PDOS, as below:
\begin{equation}
M_{\mu}(n, \mathbf{k}) = \sum_{\nu} C_{n\nu}^{*}(\mathbf{k}) S_{\nu\mu}(\mathbf{k}) C_{n\mu}(\mathbf{k}),
\end{equation}
where $M_{\mu}(n, \mathbf{k})$ is the contribution of atomic orbital $\mu$ to the energy band $n$ at the $\mathbf{k}$ point.

\subsection{Spin texture}

Spin texture refers to the spatial distribution of electron spins in momentum space, which can be measured using various techniques such as angle-resolved photoemission spectroscopy (ARPES) or scanning tunneling microscopy (STM).  In materials with spin-orbit coupling, the electron spins can be coupled to their momenta, resulting in non-trivial spin textures that can give rise to interesting physical phenomena, such as the spin Hall effect, topological insulators, and magnetic skyrmions.
In PYATB, the spin texture is calculated as follows,
\begin{equation}
\langle \Psi_{n\mathbf{k}} | \hat{\sigma}_i  | \Psi_{n\mathbf{k}} \rangle
= \sum_{\mu,\nu,s,s^{\prime}} C^{*}_{n,\mu s}(\mathbf{k}) S_{\mu\nu, s s^{\prime}}(\mathbf{k}) \hat{\sigma}_{i,s s^{\prime}} C_{n,\nu s^{\prime}}(\mathbf{k}) \, ,
\end{equation}
where $\hat{\sigma}_i$ are the Pauli matrices, with $i$= $x$, $y$, $z$, and $s$=$\uparrow$, $\downarrow$ is the spin index.

\subsection{Band unfolding}

In first-principles calculations of imperfect crystals containing disorders, defects, dopants, or alloyed atoms, supercell approximations are often used. These systems can be considered as perturbations to the original crystal structure, which break the translation symmetry of the original unit cell and introduce coupling between different $\mathbf{k}$ points in the BZ. The band structure of the supercell is folded heavily in the first BZ, making it difficult to analyze and unsuitable for comparison with angle-resolved photoemission spectroscopy (ARPES) experiments \cite{dargam_disorder_1997, wang_majority_1998}.
The band unfolding method is a powerful tool for analyzing the band structures of the supercell by projecting the Bloch wave functions of the supercell onto the coupled $\mathbf{k}$ points in the original unit cell \cite{ku_unfolding_2010, popescu_extracting_2012, lee_unfolding_2013, chen_layer_2018}. The unfolded spectra can then be directly compared with ARPES experiments.

PYATB implements the band unfolding method developed in  Ref.~\cite{dai_first-principles_2022}.
Suppose, the relationships between lattice vectors of the supercell, which denote as the large cell (LC), $\mathbf{A}$ and
the projected cell (PC) $\mathbf{a}$ are given by
\begin{equation}
\begin{bmatrix}
A_1 \\
A_1 \\
A_3
\end{bmatrix} =
\begin{bmatrix}
m_{11} & m_{12} & m_{13} \\
m_{21} & m_{22} & m_{23} \\
m_{31} & m_{32} & m_{33} \\
\end{bmatrix}
\begin{bmatrix}
a_1 \\
a_2 \\
a_3
\end{bmatrix}
,\quad m_{ij} \in \mathbb{Z} \,,
\end{equation}
or in short, $\mathbf{A} = \mathbf{M} \cdot \mathbf{a}$.
We project the Bloch wave functions of the LC, which is represented by the NAO bases, to the PW bases of the PC and therefore do not need to assume the crystal structures of PC. The spectral weight of the energy band at $\mathbf{k}_p$ can be calculated as,
\begin{equation}
A(\mathbf{k}_p, E) = \sum_{N, \mathbf{g}} |D_{N}(\mathbf{k}_p, \mathbf{g}))|^2\delta(E_N-E) \, ,
\end{equation}
where,
\begin{equation}
D_{N}(\mathbf{k}_p, \mathbf{g}) = \sum_{\mu, i} \phi_{\mu}(\mathbf{k}_p + \mathbf{g}) S_{a, i}(\mathbf{k}_p + \mathbf{g}) C_{N\mu,i}(\mathbf{K}) \, .
\label{eq:DN}
\end{equation}
In the above equation, $C_{N\mu,i}(\boldsymbol{K})$ is the eigenvector of the $N$-th band of the LC, and
\begin{eqnarray}
\phi_{\mu}(\boldsymbol{q}) &=& {1 \over \sqrt{V}} \int d\boldsymbol{r} \,\phi_{\mu}(\boldsymbol{r}) e^{-i \boldsymbol{q}\cdot \boldsymbol{r}} \, ,\\
S_{\alpha, i}(\boldsymbol{q}) &= & e^{-i \boldsymbol{q} \cdot \tau_{\alpha, i}} \, .
\end{eqnarray}
$\phi_{\mu}(\boldsymbol{q})$, which is known as the form factor of the orbital, is determined solely by the shape of the orbital,
whereas the structure information is contained in $S_{\alpha, i} (\boldsymbol{q})$.
Equation (\ref{eq:DN}) can be computed efficiently, as the number of $\phi_{\mu}(\boldsymbol{q})$ is restricted to the types of NAOs in the LC, and the number of $ \boldsymbol{g}$ vectors is determined by the size of the PC, which is typically very small. To calculate the unfolded band spectral, the energy cutoff for the $ \boldsymbol{g}$  vectors can be significantly lower than that used for self-consistent and band structure calculations.

\subsection{Berry phase and Wilson loop}

A Berry phase~\cite{zak_berrys_1989, xiao_berry_2010} is a geometric phase that describes the accumulation of phases as a wave function evolves along a closed loop in the external parameter space. In condensed matter physics, the parameter space is generally taken to be the $\mathbf{k}$-space.
The Berry phase of the $n$-th band is given by,
\begin{equation}
\phi_n = \oint_{\mathcal{C}} \mathbf{A}_{nn}(\mathbf{k}) \cdot d\mathbf{k}\,,
\end{equation}
where $\mathbf{A}_{nm}(\mathbf{k})$ is the multi-bands Berry connection, defined as,
\begin{equation}
\mathbf{A}_{n m}(\mathbf{k})=i \left\langle u_{n\mathbf{k}} | \nabla_{\mathbf{k}} | u_{m\mathbf{k}}\right\rangle\,.
\label{eq:berryconnection}
\end{equation}
The total Berry phase of a group of bands can be calculated as,
\begin{equation}
\phi = \oint_{\mathcal{C}} \mathrm{Tr}\left[\mathbf{A}\right] \cdot d\mathbf{k}\,.
\end{equation}

To calculate the Berry phase, we integrate the Berry connection on the discrete  $\mathbf{k}$-points,
using the algorithm developed in Ref.~\cite{king_theory_1993}, which gives:
\begin{equation}
\phi = -\mathrm{Im} \ln \mathrm{det} \prod_{i= 0}^{N-1} M^{(\mathbf{k}_i, \mathbf{k}_{i+1})}\,,
\end{equation}
where the overlap matrix $M^{(\mathbf{k}_i, \mathbf{k}_{i+1})}_{nm} = \langle u_{n\mathbf{k}_i} | u_{m\mathbf{k}_{i+1}}\rangle$. On the NAO base, the $M^{(\mathbf{k}_i, \mathbf{k}_{i+1})}_{nm}$ matrix is calculated as follows,
\begin{equation}
\langle u_{n\mathbf{k}_i} | u_{m\mathbf{k}_{i+1}} \rangle
= \sum_{\nu\mu} C_{n\nu}(\mathbf{k}_i)^{*} C_{m\mu}(\mathbf{k}_{i+1}) \sum_{\mathbf{R}} \mathrm{e}^{i\left(\mathbf{k}_i+\Delta\mathbf{k}\right)\cdot\mathbf{R}} \langle \mathbf{0}\nu | \mathrm{e}^{-i\Delta\mathbf{k}\cdot\mathbf{r}} | \mathbf{R}\mu \rangle
\end{equation}
where $\Delta\mathbf{k} = \mathbf{k}_{i+1} - \mathbf{k}_i$.
When $\Delta\mathbf{k}$ is very small, we can use the approximation,
\begin{equation}
\mathrm{e}^{-i\Delta\mathbf{k}\cdot\mathbf{r}} \approx 1 - i\Delta\mathbf{k}\cdot\mathbf{r}\,.
\end{equation}
To make a better approximation, we place the origin point at the midpoint between the centers of the two orbitals, namely $\frac{\mathbf{\tau}_\nu+\mathbf{\tau}_\mu+\mathbf{R}}{2}$. The overlap matrix is then calculated as,
\begin{eqnarray}
\langle \mathbf{0}\nu | \mathrm{e}^{-i\Delta\mathbf{k}\cdot\mathbf{r}} | \mathbf{R}\mu \rangle
&=& \mathrm{e}^{-i\Delta\mathbf{k}\cdot\frac{\mathbf{\tau}_\nu+\mathbf{\tau}_\mu+\mathbf{R}}{2}}
\langle \mathbf{0}\nu | \mathrm{e}^{-i\Delta\mathbf{k}\cdot\left(\mathbf{r}-\frac{\mathbf{\tau}_\nu+\mathbf{\tau}_\mu+\mathbf{R}}{2}\right)}
| \mathbf{R}\mu \rangle \nonumber \\
&=& \mathrm{e}^{-i\Delta\mathbf{k}\cdot\frac{\mathbf{\tau}_{\nu} + \mathbf{\tau}_{\mu} + \mathbf{R}}{2}} \times \left[
S_{\nu\mu}(\mathbf{R})\left(1 + i\Delta\mathbf{k}\cdot\frac{\mathbf{\tau}_{\nu} + \mathbf{\tau}_{\mu} + \mathbf{R}}{2}\right) - i\Delta\mathbf{k}\cdot\mathbf{r}_{\nu\mu}(\mathbf{R})
\right]\,.
\end{eqnarray}

The Wilson loop~\cite{yu_equivalent_2011,soluyanov_computing_2011}
is implemented in a way similar to that of the Berry phase, i.e.,
\begin{equation}
W_n(k_x) = \frac{i}{2\pi} \int_{0}^{2\pi} dk_y \, \langle u_{n, k_x, k_y} | \partial_{k_y} | u_{n, k_x, k_y} \rangle = -\frac{1}{2\pi} \mathrm{Im} \ln \prod_{i=0}^{N-1} M^{(\mathbf{k}_i, \mathbf{k}_{i+1})}_{nn} \,,
\end{equation}
where $W_n(k_x)$ is known as the Wannier charge centers (WCCs)~\cite{soluyanov_computing_2011}. The WCCs is obtained by a parallel-transport construction using $M^{(\mathbf{k}_i, \mathbf{k}_{i+1})}_{nn}$.
To achieve optimal alignment between the states of two $\mathbf{k}$ points, we construct the ``unitary part'' of $M$ to obtain $W_n(k_x)$~\cite{wu_wannier-based_2006}. We perform the singular-value decomposition (SVD) on $M = V\Sigma W^{\dagger}$, where $V$ and $W$ are unitary, and $\Sigma$ is approximately unitary. We define $\tilde{M} = VW^{\dagger}$, which is a unitary matrix, and $\Lambda = \prod_{i=0}^{N-1} \tilde{M}^{(\mathbf{k}_i, \mathbf{k}_{i+1})}$, which is also unitary.
 The eigenvalues $\lambda_n$ of $\Lambda$ are all unimodular and
 the WCCs can be expressed in terms of these eigenvalues $\lambda_n$
 ~\cite{wu_wannier-based_2006, soluyanov_wannier_2011},
\begin{equation}
W_n(k_x) =  -\frac{1}{2\pi} \mathrm{Im} \ln \lambda_n \,.
\end{equation}

\subsection{Berry curvature}

Berry curvature~\cite{thouless_quantized_1982, chang_berry_2008} plays an essential role in descirbing the topological properties of energy bands and the dynamics of Bloch electron.
The calculation of Berry curvature on NAO bases has been given in Ref.~\cite{jin_calculation_2021}.

The multi-bands Berry curvature is defined as,
\begin{equation}
{\Omega}_{nm,ab}(\mathbf{k})
= \partial_a\mathbf{A}_{nm,b}(\mathbf{k}) - \partial_b\mathbf{A}_{nm,a}(\mathbf{k})
= i\langle\partial_a u_{n\mathbf{k}}|\partial_b u_{m\mathbf{k}}\rangle - i\langle\partial_b u_{n\mathbf{k}}|\partial_a u_{m\mathbf{k}}\rangle\, ,
\label{eq:omega_m}
\end{equation}
where $\partial_{a} = \partial / \partial_{k_{a}}$, $a = x, y, z$. Substituting Eq.~(\ref{eq:cell_periodic_wf}) into Eq.~(\ref{eq:omega_m}), we have
\begin{eqnarray}
i\langle\partial_a u_{n\mathbf{k}}|\partial_b u_{m\mathbf{k}}\rangle
&=&i\sum_{\nu,\mu}C_{n\nu}^{*}(\mathbf{k})C_{m\mu}(\mathbf{k})\sum_{\vec{R}}\mathrm{e}^{i\mathbf{k}\cdot\mathbf{R}}\langle\mathbf{0}\nu|-r_a(R_b-r_b)|\mathbf{R}\mu\rangle \nonumber \\
&+& i\sum_{\nu,\mu}\left(\partial_a C_{n\nu}^{*}(\mathbf{k})\right)S_{\nu\mu}(\mathbf{k})\left(\partial_b C_{m\mu}(\mathbf{k})\right) \nonumber \\
&+& \sum_{\nu,\mu}\left(\partial_a C_{n\nu}^{*}(\mathbf{k})\right)C_{m\mu}(\mathbf{k})\sum_{\mathbf{R}}\mathrm{e}^{i\mathbf{k}\cdot\mathbf{R}}\langle0\nu|r_b-R_b|\mathbf{R}\mu\rangle \nonumber \\
&-& \sum_{\nu,\mu}C_{n\nu}^{*}(\mathbf{k})\left(\partial_b C_{m\mu}(\mathbf{k})\right)\sum_{\mathbf{R}}\mathrm{e}^{i\mathbf{k}\cdot\mathbf{R}}\langle0\nu|r_a|\mathbf{R}\mu\rangle \label{eq:uu_eq}
\end{eqnarray}
Then, we can simplify Eq.~(\ref{eq:uu_eq}) by inserting the identity matrix,
\begin{equation}
I = \sum_{n}C_{n}(\mathbf{k})C_{n}^\dagger(\mathbf{k}) S(\mathbf{k}) = \sum_{n}S(\mathbf{k})C_n(\mathbf{k})C_n^\dagger(\mathbf{k}) \, .
\end{equation}
and introduce the relevant definitions,
\begin{eqnarray}
\bar{A}_{nm,a}(\mathbf{k}) &=& C_n^\dagger(\mathbf{k}) A^R_a(\mathbf{k}) C_m(\mathbf{k}) \, , \\
D_{nm,a}(\mathbf{k}) &=& C_n^\dagger(\mathbf{k}) S(\mathbf{k})\left(\partial_a C_m(\mathbf{k})\right) \, ,
\end{eqnarray}
Eventually, after some algebraic operations, we obtain the Berry curvature under NAOs~\cite{jin_calculation_2021},
\begin{equation}
\Omega_{nm,ab} = \bar{\Omega}_{nm,ab}
+ i\left(D^\dagger_a D_b-D^\dagger_b D_a\right)_{nm}
+ \left(D^\dagger_a\bar{A}^\dagger_b+\bar{A}_b  D_a\right)_{nm} -\left(D^\dagger_b\bar{A}^\dagger_a+\bar{A}_a D_b\right)_{nm} \, ,
\label{eq:origin_curv}
\end{equation}
where
\begin{equation}
\label{eq:omega_bar}
\bar{\Omega}_{nm,ab}(\mathbf{k}) = i\sum_{\nu,\mu}C_{n\nu}^{*}(\mathbf{k})C_{m\mu}(\mathbf{k})\sum_{\mathbf{R}}\mathrm{e}^{i\mathbf{k}\cdot\mathbf{R}}\langle0\nu|r_b R_a-r_a R_b|\mathbf{R}\mu\rangle \, .
\end{equation}
For both $\bar{\Omega}_{nm,ab}(\mathbf{k})$ and $\bar{A}_{nm,a}(\mathbf{k})$ matrices are directly obtainable from the tight binding model, while $D_{nm,a}(\mathbf{k})$ matrix is calculated by linear response theory,
\begin{equation}
D_{nm,a}(\mathbf{k}) =
\frac{\bar{H}_{nm,a}(\mathbf{k})-E_{m\mathbf{k}} \bar{S}_{nm,a}(\mathbf{k})}{E_{m\mathbf{k}}-E_{n\mathbf{k}}} \quad (n\neq m) \, .
\label{eq:d-matrix}
\end{equation}
where
\begin{eqnarray}
\bar{H}_{nm,a}(\mathbf{k}) &=& C_n^\dagger(\mathbf{k})\left(\partial_a H(\mathbf{k})\right)C_m(\mathbf{k}) \, , \\
\bar{S}_{nm,a}(\mathbf{k}) &=& C_n^\dagger(\mathbf{k})\left(\partial_a S(\mathbf{k})\right)C_m(\mathbf{k}) \, .
\end{eqnarray}

The total Berry curvature is defined as,
\begin{equation}
\Omega_{ab}(\mathbf{k})= \mathrm{Tr}\left[\Omega_{nm,ab}(\mathbf{k})\right]  = \sum_{n}f_n(\mathbf{k})\Omega_{nn,ab}(\mathbf{k})\, ,
\label{eq:total_bc}
\end{equation}
where $\mathrm{Tr}$ denotes a trace over the occupied bands,  and $f_n(\mathbf{k})$ is the Fermi occupation function. Note that the trace of a multi-bands Berry curvature and of a non-Abelian Berry curvature is the same.

The Berry curvature can also be calculated by Kubo formula,
\begin{equation}
\Omega_{ab}^{\text{Kubo}}(\mathbf{k}) = -2 \operatorname{Im} \sum_{n}^{\text{occ}} \sum_{m}^{\text{uocc}} \frac{v_{nm,a}(\mathbf{k}) v_{mn,b}(\mathbf{k})}{\left(E_{m\mathbf{k}} - E_{n\mathbf{k}}\right)^2} \, ,
\label{eq:bc_kubo}
\end{equation}
where $v_{nm, a}(\mathbf{k})$ is the velocity matrix. The velocity matrix is a fundamental physical quantity in the optical response of solid materials and is also closely linked to the Berry connection, which is given by,
\begin{equation}
v_{nm,a}(\mathbf{k}) = \langle \Psi_{n\mathbf{k}} | \hat{v}_{a} | \Psi_{m\mathbf{k}} \rangle = \left(\partial_a E_{n\mathbf{k}}\right) \delta_{nm}
- i\left(E_{m\mathbf{k}}-E_{n\mathbf{k}}\right)A_{nm,a}(\mathbf{k})\, ,
\end{equation}
where the multi-bands Berry connection under the NAOs is in the form:
\begin{equation}
\label{eq:bc_nao}
A_{nm, a}(\mathbf{k})=i D_{nm, a}(\mathbf{k}) + \bar{A}^\dagger_{nm, a}(\mathbf{k})\,.
\end{equation}
After some derivation, we can obtain the expression for the velocity matrix on the NAO base as follows,
\begin{equation}
\upsilon_{nm,a}(\mathbf{k}) = \bar{H}_{nm,a}(\mathbf{k}) - E_{n\mathbf{k}}\bar{S}_{nm,a}(\mathbf{k}) + i(E_{n\mathbf{k}}-E_{m\mathbf{k}})\bar{A}_{nm,a}(\mathbf{k}) \,.
\end{equation}
This formula of velocity matrix can also be derived from $\hat{\mathbf{v}} \equiv \dot{\mathbf{r}} = \left(i/\hbar\right)[\hat{H}, \mathbf{r}]$ under non-orthogonal NAOs~\cite{lee_tight-binding_2018}.

The Berry curvature of Eq.~(\ref{eq:bc_kubo}) has small difference to that of Eq.~(\ref{eq:total_bc}), which is due to the incompleteness of the NAO base to the original Hilbert space. However, it has been shown that the correction terms are usually very small even for the double-$\zeta$ plus polarization (DZP) basis set~\cite{jin_calculation_2021}.  PYATB has implemented both formulations, and the default method for computing the Berry curvature is Eq.~(\ref{eq:total_bc}).

The Chern number is a topological invariant used to classify topological materials. It is closely related to the quantum Hall effect and the quantum anomalous Hall effect. The Chern number is obtained by integrating the Berry curvature on any closed 2D manifold, i.e.,
\begin{equation}
C = \frac{1}{2\pi} \oint_{\mathbf{S}} \Omega \cdot d\mathbf{S} \,.
\end{equation}

\subsection{Optical conductivity and dielectric functions}

JDOS determines the number of permissible optical transitions from the valence bands to the conduction bands at a particular energy, and is intimately linked to the dielectric function and optical conductivity, which is given by
\begin{equation}
D_{\text{joint }}(\omega)=\frac{V_{\text{cell}}}{\hbar} \int \frac{d^3k}{(2\pi)^3} \sum_{n, m} f_{n m} \delta\left(\omega_{m n}-\omega\right)\,.
\end{equation}
The optical properties of semiconductor materials are characterized by the dielectric function, optical conductivity, and absorption coefficient, etc.

The Kubo-Greenwood formula, based on the independent-particle approximation, are used to calculate the optical conductivity and dielectric function using the velocity matrix,
\begin{equation}
\sigma_{ab}(\omega) = \frac{i e^2\hbar}{N_k V_{\text{cell}}}\sum_{\mathbf{k}}
\sum_{n,m}\left(\frac{f_{mn}}{\omega_{mn}}\right)
\frac{v_{nm, a} v_{mn, b}}{\hbar\omega_{mn} - \left(\hbar\omega + i\eta\right)},
\end{equation}
where $N_k$ is the number of $\mathbf{k}$ points, and $V_{\text{cell}}$ is the cell volume. $f_{nm}$=$f_n-f_m$ and $\hbar \omega_{nm}$=$E_n-E_m$ are differences between Fermi occupation factors and band energies, respectively.

The imaginary part of the dielectric function is calculated from the following equation,
\begin{equation}
\epsilon_i^{ab}(\omega) = -\frac{e^2 \pi}{\epsilon_0 \hbar N V_{\text{cell}}} \sum_{\mathbf{k}} \sum_{n,m} f_{nm} \frac{v_{nm, a} v_{mn, b}}{\omega_{mn}^2} \delta\left(\omega_{mn} - \omega\right),
\end{equation}
and the real part of the dielectric function is obtained by the Kramer-Kronig transformation,
\begin{equation}
\epsilon_{r}^{ab}(\omega) = \delta_{ab} + \frac{2}{\pi} \mathbf{P} \int_{0}^{\infty} d\omega^{\prime} \frac{\omega^{\prime}\epsilon_{i}^{ab}\left(\omega^{\prime}\right)}{\omega^{\prime 2} - \omega^2},
\end{equation}
where $\mathbf{P}$ denotes the principal value of the integral.
The absorption coefficient $\alpha(\omega)$ can be calculated from dielectric functions,
\begin{equation}
\alpha(\omega) = \frac{\sqrt{2} \omega}{c} \left(\sqrt{\epsilon_r^2+\varepsilon_i^2}-\epsilon_r\right)^{\frac{1}{2}}\,.
\label{eq:absorb}
\end{equation}

\subsection{Shift current conductivity}

The shift current is an intrinsic contribution to the bulk photovoltaic effect (BPVE)~\cite{sturman_photovoltaic_1992, sipe_second-order_2000}. It describes the photocurrent generated by light illumination on homogeneous non-centrosymmetric crystals.
The shift current is a second-order optical response. It can be expressed as a DC current, generated by a monochromatic photoelectric field $\mathbf{E}(t) = \mathbf{E}(\omega)\mathrm{e}^{i\omega t} + \mathbf{E}(-\omega)\mathrm{e}^{-i\omega t}$, where
\begin{equation}
J^{a} = 2 \sigma^{abc}(0; \omega, -\omega) E_{b}(\omega) E_{c}(-\omega).
\end{equation}
Here, $a, b, c = x, y, z$, and $\sigma^{abc}(0; \omega, -\omega)$ is the shift current tensor,
\begin{equation}
\label{eq:shift_current}
\sigma^{abc}(0 ; \omega,-\omega) =	\frac{\pi e^3}{\hbar^2} \int \frac{d \mathbf{k}}{8 \pi^3} \sum_{n, m} f_{nm} \mathrm{Im}\left[ I_{mn}^{abc} + I_{mn}^{acb} \right] \delta\left(\omega_{m n}-\omega\right) \,,
\end{equation}
where $I_{mn}^{abc} = r_{mn}^{b} r_{nm;a}^{c}$, $r_{nm}^a$ is the inter-band dipole matrix, and$r_{nm; a}^b$ is the generalized derivative of the dipole matrix, i.e.,
\begin{eqnarray}
r_{nm}^{a}  &=& (1-\delta_{nm}) A_{nm, a}, \label{eq:dipole matrix}\\
r_{nm;b}^{a} &=& \partial_{b} r_{nm}^{a} - i\left(A_{nn, b} - A_{mm, b}\right) r_{nm}^{a} . \label{eq:dr_origin}
\end{eqnarray}

The inter-band dipole matrix $r_{nm}^{a}$ can be obtained using Eq.~(\ref{eq:bc_nao}).
To calculate $r_{nm; a}^b$, we substitute Eq.~(\ref{eq:dipole matrix}) into Eq.~(\ref{eq:dr_origin}), which gives us
\begin{equation}
r_{nm;b}^{a} = i\partial_{b}D_{nm,a} + \partial_{b}\bar{A}_{nm,a}^{\dagger} - i \left(
i D_{nn,b} + \bar{A}_{nn,b}^{\dagger} - i D_{mm,b} - \bar{A}_{mm,b}^{\dagger}
\right) \left(\ i D_{nm,a} + \bar{A}_{nm,a}^{\dagger} \right).
\end{equation}
Following the derivation of Ref.~\cite{ibanez-azpiroz_ab_2018}, the general matrix containing k takes the following form,
\begin{equation}
\bar{\mathcal{O}} = C^\dagger \mathcal{O} C
\end{equation}
and using the identity
\begin{equation}
\partial_{a} C = C D_{a}
\end{equation}
we have
\begin{equation}
\partial_{a} \bar{\mathcal{O}} = D_a^\dagger \bar{\mathcal{O}} + \bar{\mathcal{O} D_a} + C^\dagger \partial_{a}\mathcal{O} C
\end{equation}
We calculate $\partial_{b} D_{nm, a}$ by applying the above equations to Eq.~(\ref{eq:d-matrix}), and
obtain the expression for $r_{nm;b}^{a}$ on the non-orthogonal atomic orbitals as follows,
\begin{eqnarray}
r_{nm;b}^{a}
&=& \frac{i}{E_m - E_n} \left[\sum_{l}D_{nl, b}^{\dagger}\bar{H}_{lm, a} + \sum_{l} \bar{H}_{nl, a}D_{lm, b} + \bar{H}_{nm, ab}\right] \nonumber \\
&-& \frac{i E_m}{E_m - E_n}\left[\sum_{l}D_{nl, b}^{\dagger}\bar{S}_{lm, a} + \sum_{l} \bar{S}_{nl, a}D_{lm, b} + \bar{S}_{nm, ab}\right] \nonumber \\
&-& \frac{i \left(\partial_{b}E_{m}\right)\bar{S}_{nm, a}}{E_m - E_n} \nonumber \\
&-& \frac{i \left(\bar{H}_{nm, a} - E_m \bar{S}_{nm, a}\right)\left(\partial_{b}E_m - \partial_{b}E_n\right)}{\left(E_m - E_n\right)^2}  \nonumber \\
&+& \left[\sum_{l}D_{nl, b}^{\dagger}\bar{A}_{lm, a} + \sum_{l} \bar{A}_{nl, a}D_{lm, b} + \bar{A}_{nm, ab}\right]
\end{eqnarray}
where
\begin{eqnarray}
\bar{H}_{nm, ab} &=& C_n^\dagger \left( \partial_{a} \partial_{b} H \right) C_m \\
\bar{S}_{nm, ab} &=& C_n^\dagger \left( \partial_{a} \partial_{b} S \right) C_m \\
\bar{A}_{nm, ab} &=& C_n^\dagger \left( \partial_{b} A_{a}^{R} \right)  C_m
\end{eqnarray}
In the derivation of the above equation, we use the parallel transport gauge, $A_{nn, a} = iD_{nn, a} + \bar{A}_{nn, a}^{\dagger} = 0$, which implies $D_{nn,a}=i \bar{A}_{nn,a}^{\dagger}$. This allows us to give the full expression of $D_{nm,a}$ as follows,
\begin{equation}
D_{nm,a} = C_n^{\dagger} S C_m^{(a)} = \left\{
\begin{aligned}
& i \bar{A}_{nm,a}^{\dagger} & \quad & n = m \\
& \frac{\bar{H}_{nm, a} - E_m \bar{S}_{nm, a}}{E_m - E_n + i\eta} & \quad & n\neq m, \eta\to 0
\end{aligned}
\right.
\end{equation}
where $\eta$ is introduced to avoid numerical problems that may arise due to nearly degenerate energy bands.

\subsection{Berry curvature dipole}

In a system with time-reversal symmetry, the Berry curvature is an odd function of $\mathbf{k}$, i.e., $\Omega_{a}(\mathbf{k})=-\Omega_{a}(-\mathbf{k})$. As a result, the integration of the Berry curvature over the BZ is zero.
However, if the system lacks
a inversion symmetry, a higher-order nonlinear AHC can arise ~\cite{sipe_second-order_2000,  sodemann_quantum_2015}.
More specifically,
$j_a^0=\chi_{abc} E_b (\omega) E_c(-\omega)$  and
$j_a^{2 \omega}=\chi_{abc} E_b (\omega)E_c(\omega)$,
describe a rectified current and the second harmonic, respectively, whereas $\omega$ is the driving frequency.
The coefficient $\chi_{abc}$ is given by
\begin{equation}
\chi_{abc}=-\varepsilon_{a d c} \frac{e^3 \tau}{2(1+i \omega \tau)} D_{bd} .
\end{equation}
where,
\begin{equation}
D_{ab}=\int_{k} f_{0}\left(\frac{\partial \Omega_{b}}{\partial k_{a}} \right)
\end{equation}
is called the Berry curvature dipole. In practice, it is more convenient to  calculate $D_{ab}$ using the following
formula~\cite{sodemann_quantum_2015}:
\begin{equation}
D_{ab}=-\int_{k}\left(\frac{\partial f_0}{\partial E} \right) \left(\frac{\partial E}{\partial k_{a}} \right)\Omega_{b} \,.
\end{equation}
The nonlinear AHC has many important applications, such as the Terahertz detection~\cite{zhang_terahertz_2021}.
Further realization demands a summation over all bands. At given temperature $T$, we have,
\begin{equation}
D_{a b}(T)=\int[d k] \sum_{n} \frac{\partial E_{n}}{\partial k_{a}} \Omega_{n,b}\left(-\frac{\partial f_{0}}{\partial E}\right)_{E=E_{n}}.
\end{equation}
However, this approach requires calculating the Berry curvature dipole at each temperature, which can be computationally demanding when a large number of temperatures are required.
Alternatively, we can first calculate  $D_{a b}(E)$ as follows~\cite{tsirkin_gyrotropic_2018},
\begin{equation}
D_{a b}(E) =  \int[d k] \sum_{n} \frac{\partial E_{n}}{\partial k_{a}} \Omega_{n,b}\delta(E_{n}-E).
\end{equation}
Then the Berry curvature dipole at given temperature $T$ and chemical potential $\mu$, can be easily calculated as,
\begin{equation}
D_{a b}(\mu, T) = -\int \frac{\partial f_{0}(E,\mu, T)}{\partial E} D_{a b}(E) dE .
\end{equation}

\section{Installation and running}
\label{sec:install}

In this section,  we present a guide on how to install and utilize PYATB. PYATB is available on the public GitHub repository at https://github.com/pyatb/pyatb. The main PYATB code is written in C++, and python extensions are provided via pybind11. Matrix calculations are performed using the Eigen library in C++, which can be accelerated by adding linear algebra libraries such as BLAS and LAPACK.
To install PYATB, you need to specify the C++ compiler and linear algebra library in the {\tt setup.py} file, and then follow the standard python software installation process by running {\tt python setup.py install}. After installation, the executable {\tt pyatb} file will be added to your python environment, and the pyatb module will be available for use.

Before running the PYATB program, four input files are required: {\tt HR}, {\tt SR}, {\tt rR}, and {\tt Input}. The first three files contain the data of the tight-binding model, including $H_{\nu\mu}(\mathbf{R})$, $S_{\nu\mu}(\mathbf{R})$, and $r_{\nu\mu,a}(\mathbf{R})$, respectively.
Some functionalities, such as {\tt band unfolding}, {\tt PDOS} and {\tt fat band}, the structure file and orbital files are also required.
The {\tt Input} file is used to specify the material structure and setup parameters for each function.
Currently, PYATB has an interface with the first-principles package ABACUS. The {\tt HR}, {\tt SR}, and {\tt rR} files can be automatically generated by performing self-consistent calculations in ABACUS. It is straightforward to develop interfaces with other NAO-based first-principles packages.

PYATB supports a mixed parallelism of MPI and OpenMP. After preparing the four input files, the program can be run as follows (for example):
\begin{verbatim}
$ export OMP_NUM_THREADS=2
$ mpirun -np 6 pyatb
\end{verbatim}
During the execution of the program, multiple output files will be created. All of these files are stored in the {\tt Out} folder.
The {\tt running.log} file keeps track of the current status of the program. The output files for each individual function are stored in their respective folders. Moreover, some functions may generate images using matplotlib.

\section{Examples}
\label{sec:examples}

In this section, we provide examples of six different physical systems to illustrate the various capabilities of PYATB. These examples include the nitrogen-vacancy (NV) center in diamond, Bi$_2$Se$_3$, MnSb$_2$Te$_4$, CsPbI$_3$, WS$_2$ and Te. For each of these examples, we first generate the necessary input files, {\tt HR}, {\tt SR}, and {\tt rR}, using ABACUS. The ABACUS input files for each example is also provided in the PYATB {\tt examples/} directory.

\subsection{Band unfolding}

\begin{figure}[!tp]
	\centering
	\includegraphics[width=0.8\textwidth]{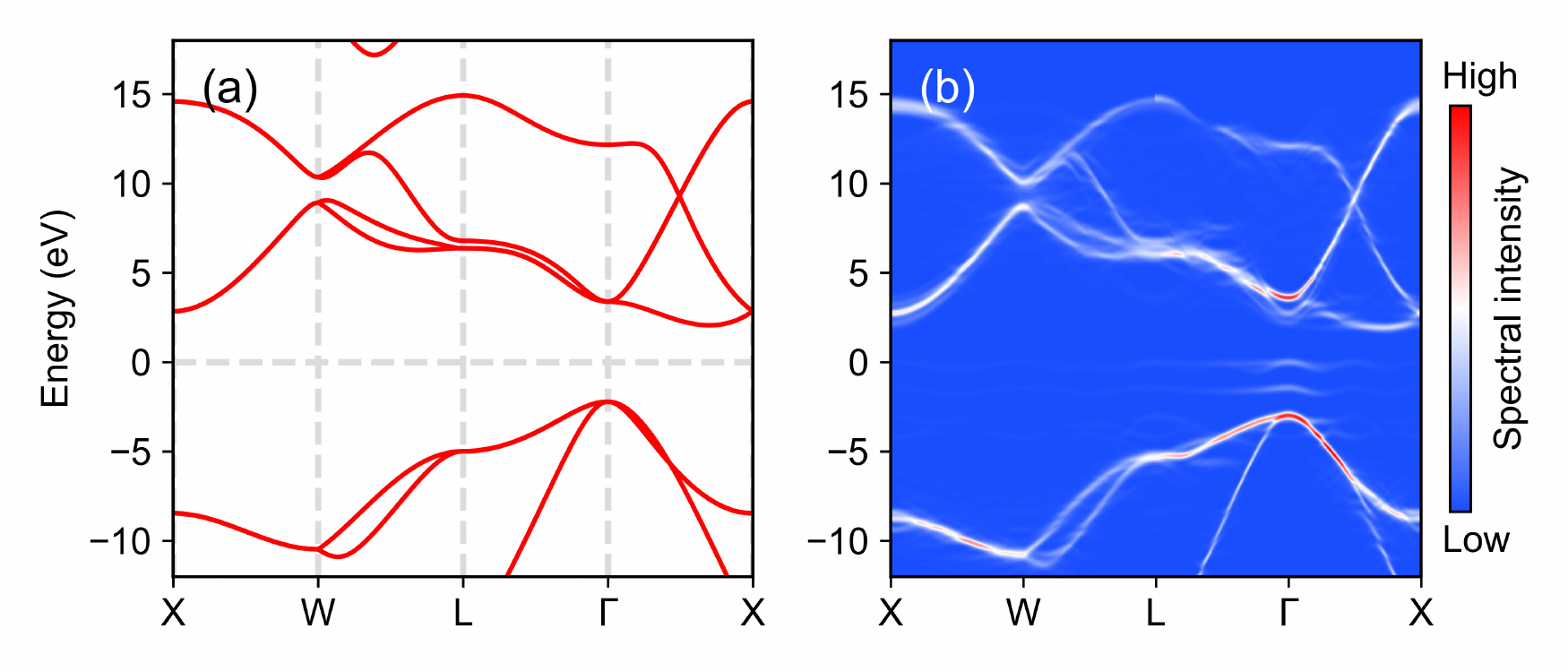}
	\caption{(a) Band structure of diamond in the primitive unit cell. (b) Unfolded band spectrum of the NV center in a supercell.}
	\label{fig:NV}
\end{figure}

The NV center is a point defect in diamond that plays a crucial role in emerging quantum technologies. In this example, we showcase PYATB's band unfolding function by calculating the spectral function of the NV center in diamond.
To construct the NV center, we replace two C atoms with a N atom and a vacancy in a supercell containing 2$\times$2$\times$2 8-atoms conventional diamond unit cells.
Figure~\ref{fig:NV}(a) shows the band structure of the diamond primitive cell, while Fig.\ref{fig:NV}(b) shows the unfolded band structure spectra obtained by the band unfolding method. As depicted in Fig.\ref{fig:NV}(b), impurity bands appear near the $\Gamma$ point.

An example {\tt Input} file for performing band unfolding calculations using PYATB is provided in \ref{input_nv}.
To do the band unfolding calculations, you also need the structure and NAO files.

\subsection{Spin texture and Wilson loop}

In this example, we demonstrate how to use PYATB to calculate the spin texture and $\mathbb{Z}_2$ number for Bi$_2$Se$_3$. The $\mathbb{Z}_2$ number is a topological invariant that characterizes whether a band insulator with time-reversal symmetry would possess topological properties. In three-dimensional (3D) systems, there are four independent $\mathbb{Z}_2$ numbers, consisting of one strong topological index and three weak topological indices. These four $\mathbb{Z}_2$ numbers enable the classification of 3D time-reversed band insulators into strong topological insulators, weak topological insulators, and trivial insulators. The Wilson loop method provides a visual means of computing the $\mathbb{Z}_2$ number.

\begin{figure*}[htbp]
	\centering
	\includegraphics[width=0.5\textwidth]{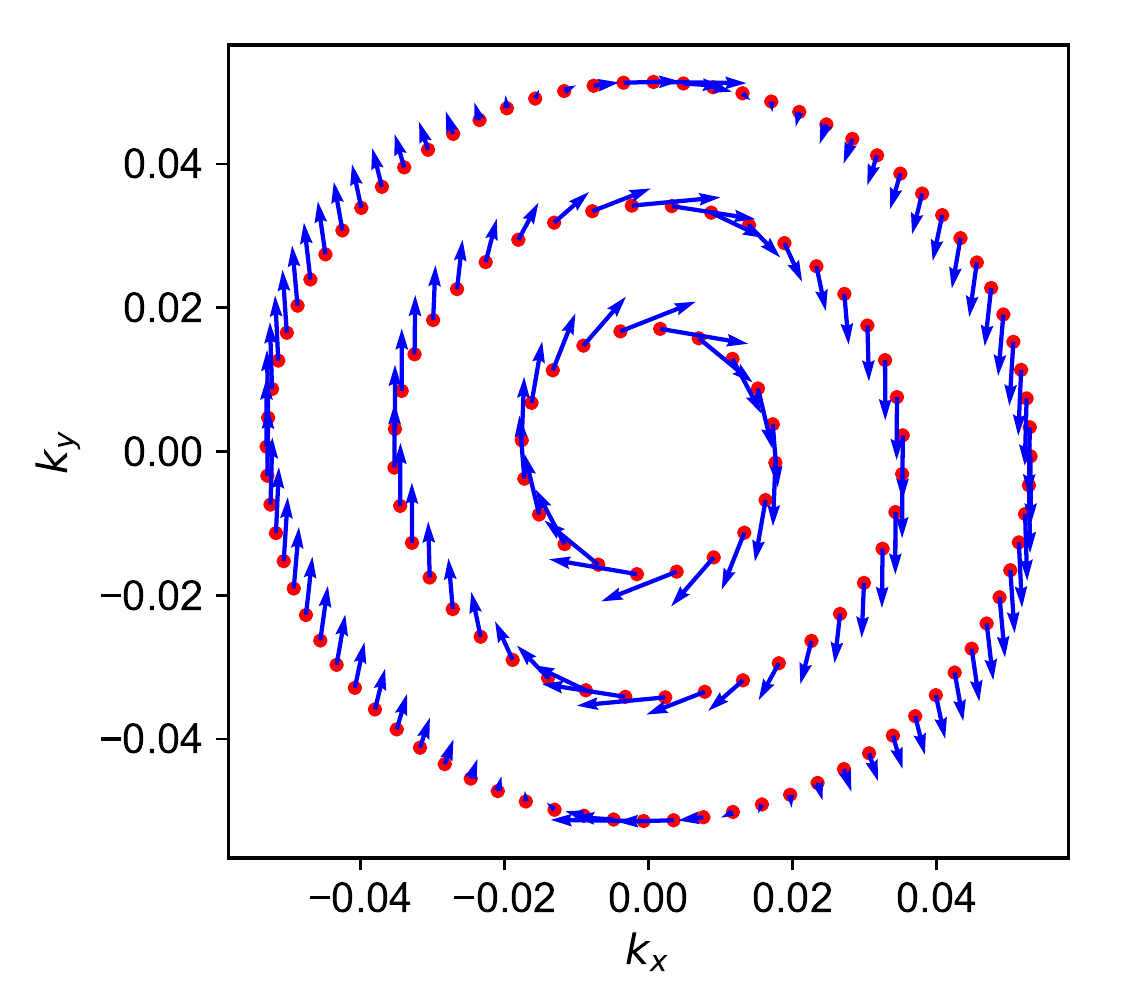}
	\caption{The figure shows the Rashba-type spin texture of the highest occupied energy band in Bi$_2$Se$_3$.}
	\label{fig:Bi2Se3_spin}
\end{figure*}

\begin{figure*}[htbp]
	\centering
	\includegraphics[width=0.7\textwidth]{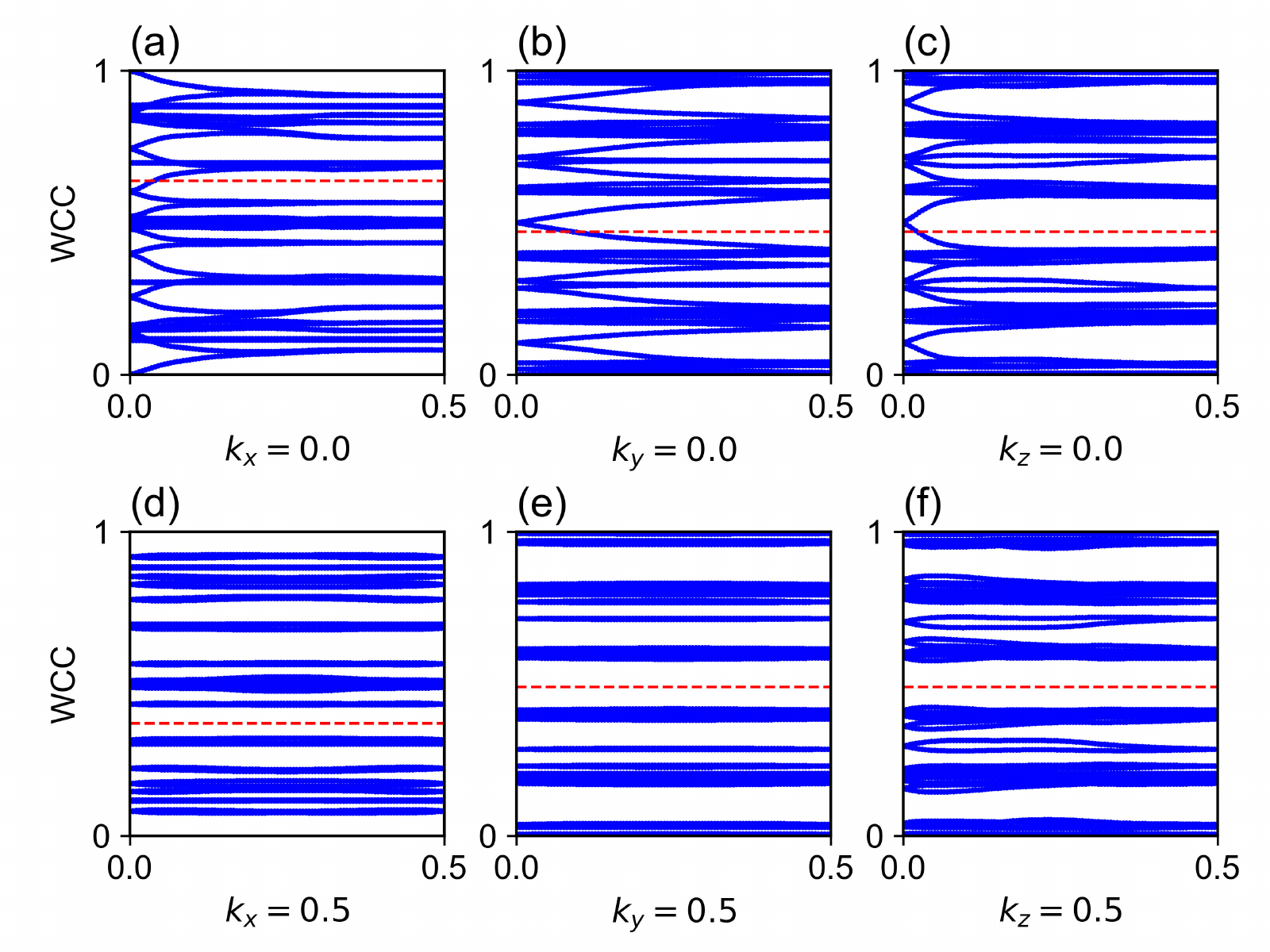}
	\caption{Wilson loops for six time-reversal invariant planes of Bi$_2$Se$_3$: (a) $k_x=0$, (b) $k_y=0$, (c) $k_z=0$, (d) $k_x=0.5$, (e) $k_y=0.5$, and (f) $k_z=0.5$. The red reference line intersects the Wilson loop an odd number of times for the first three planes, indicating $\mathbb{Z}_2=1$, while there is no intersection for the other three planes, indicating $\mathbb{Z}_2=0$.
}
	\label{fig:Bi2Se3}
\end{figure*}

Figure~\ref{fig:Bi2Se3_spin} depicts the spin texture of Bi$_2$Se$_3$ in the $k_x$-$k_y$ plane for the highest occupied energy band, which is of Rashba-type.
Figure~\ref{fig:Bi2Se3} shows the Wilson loops of Bi$_2$Se$_3$ of the six time-reversal invariant planes in the BZ, which can be used to determine its topological indices ($\nu_0, \nu_1\nu_2\nu_3$).
The results show that the red reference line intersects the Wilson loop an odd number of times at the planes $k_x=0$, $k_y=0$, and $k_z=0$, indicating that the $\mathbb{Z}_2$ index is 1. Conversely, for the planes $k_x=0.5$, $k_y=0.5$, and $k_z=0.5$, there is no intersection between the reference line and the Wilson loop, indicating that the $\mathbb{Z}_2$ index is 0. Based on these results, we obtain the $\mathbb{Z}_2$ topological indices for Bi$_2$Se$_3$ as (1,000), confirming that it is a strong topological insulator.

An example {\tt Input} file for calculating spin textures and Wilson loops using PYATB is provided in \ref{input_bi2se3}.

\begin{figure*}[htbp]
	\centering
	\includegraphics[width=0.5\textwidth]{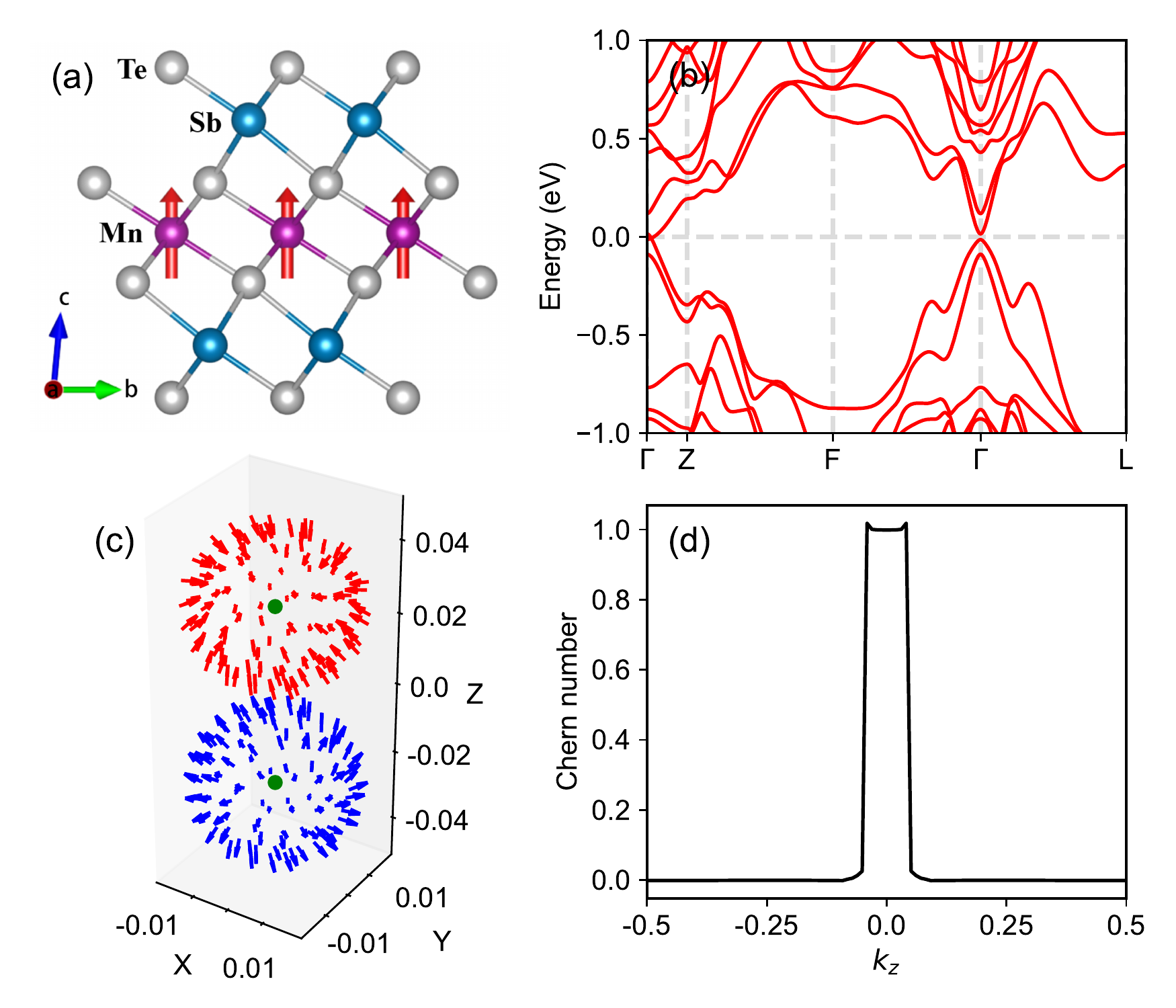}
	\caption{(a) Crystal structure of MnSb$_2$Te$_4$. (b) Band structure of MnSb$_2$Te$_4$ under FM magnetization.
A Weyl point is observed near the Fermi energy on the high symmetry line along $\Gamma-Z$. (c) Two Weyl points in the BZ, as well as the Berry curvature on the spheres surrounding them. The green dots represent the Weyl points, whereas the red and blue arrows indicate the directions of Berry curvature. (d) The Chern number of the $k_x-k_y$ plane calculated at different $k_z$.
}
	\label{fig:MnSbTe}
\end{figure*}

\subsection{Berry curvature, Chern number, Chirality}

MnSb$_2$Te$_4$ is a magnetic topological insulator in the antiferromagnetic (AFM) state, but
becomes a Weyl semimetal in the ferromagnetic (FM) state~\cite{yan_evolution_2019, wimmer_mnrich_2021, shi_anomalous_2020}.
We use the Weyl semimetal state of MnSb$_2$Te$_4$ to showcase the capabilities of PYATB, including finding nodes, calculating Berry curvature, Chern number, and chirality functions.

The unit cell of MnSb$_2$Te$_4$, consist of septuple layers (SL), is shown in Fig.~\ref{fig:MnSbTe}(a) with all spins in Mn atoms aligned in parallel. Figure~\ref{fig:MnSbTe}(b) depicts the band structures of MnSb$_2$Te$_4$ in the FM state.
A band crossing point near the Fermi energy is observed on the $\Gamma-Z$ high symmetry line, which corresponds to a Weyl point. We utilized the find nodes function, and find two band crossing points in the BZ near the Fermi energy. To confirm that they are indeed Weyl points, we calculated the chirality  of these two points, and the results are shown in Fig.~\ref{fig:MnSbTe}(c).
The green dots in Fig.~\ref{fig:MnSbTe}(c) represent a pair of Weyl points in the BZ, while the red and blue arrows indicate the direction of Berry curvature on the spheres around the Weyl points. At the Weyl point with $k_z > 0$, the Berry curvature points inward (red arrows), and the integral over the sphere gives the Chern number (chirality) equal to -1.
For the Weyl point with $k_z < 0$, the Berry curvature points outward (blue arrows), and the chirality equals +1.
We further calculated the Chern number of the $k_x-k_y$ plane at $k_z$ and find that the Chern number is 1 between the two Weyl points, and changes abruptly from 1 to 0 upon crossing the Weyl point, as shown in Fig.~\ref{fig:MnSbTe}(d).

An example {\tt Input} file for performing the above-mentioned calculations in PYATB is provided in \ref{input_mnsb2te4}.

\subsection{JDOS, dielectric function}

\begin{figure*}[htbp]
	\centering
	\includegraphics[width=0.8\textwidth]{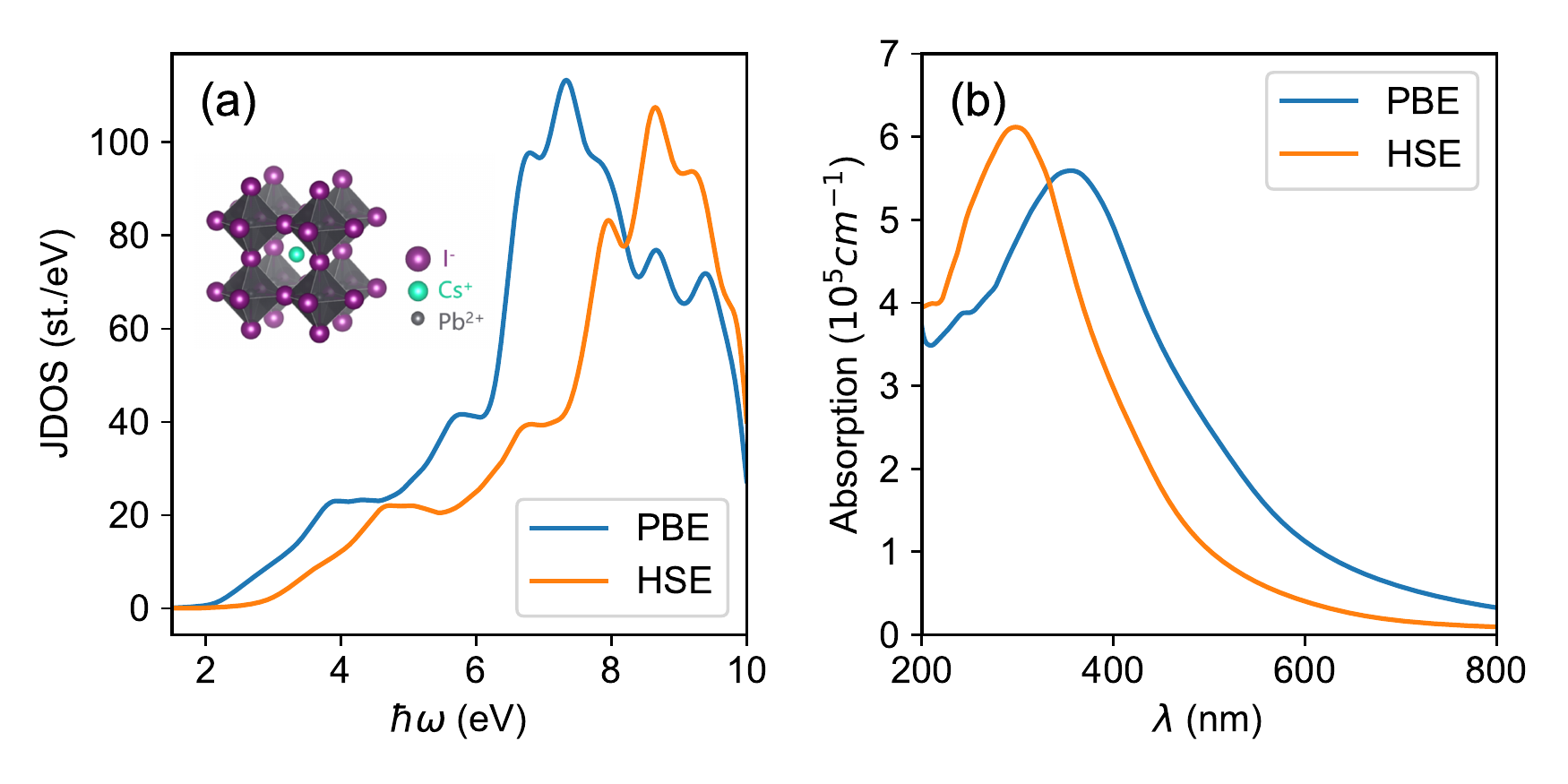}
	\caption{(a) Joint density of states and (b) absorption coefficient of CsPbI$_3$. The inset figure in (a) depicts the crystal structure of CsPbI$_3$.}
	\label{fig:CsPbI3}
\end{figure*}

In this example, we calculate the JDOS and optical absorption coefficient of CsPbI$_3$ using PYATB. CsPbI$_3$ is an all-inorganic halide perovskite and is considered one of the most promising photovoltaic materials due to its exceptional optoelectronic properties, including a long carrier diffusion length and high photoluminescence (PL) quantum yields~\cite{dastidar_slow_2017, jing_surface_2017}.
To compare the results, we employed both the Perdew-Burke-Ernzerhof (PBE)\cite{perdew_generalized_1996} functional and the Heyd-Scuseria-Ernzerhof (HSE)\cite{heyd_hybrid_2003, heyd_erratum_2006, krukau_influence_2006} hybrid functional in the calculations.
The peak position of JDOS in Fig.\ref{fig:CsPbI3}(a) obtained using HSE is shifted to the right relative to the PBE peak since PBE tends to underestimate the band gap, while HSE yields more reasonable results. The optical absorption coefficient obtained from HSE also exhibits a blue-shifted onset compared to the PBE one, as shown in Fig.\ref{fig:CsPbI3}(b). Furthermore, the HSE spectrum exhibits a strong absorption centered at $\sim$300 nm, consistent with previous theoretical predictions~\cite{liu_insight_2020}.

An example {\tt Input} file for calculating the joint density of states (JDOS) and optical conductivity is provided in \ref{input_cspbi3}, and the absorption coefficient is calculated using Eq.~(\ref{eq:absorb}).

\subsection{Fat band and shift current}

\begin{figure}[!tp]
	\centering
	\includegraphics[width=0.8\linewidth]{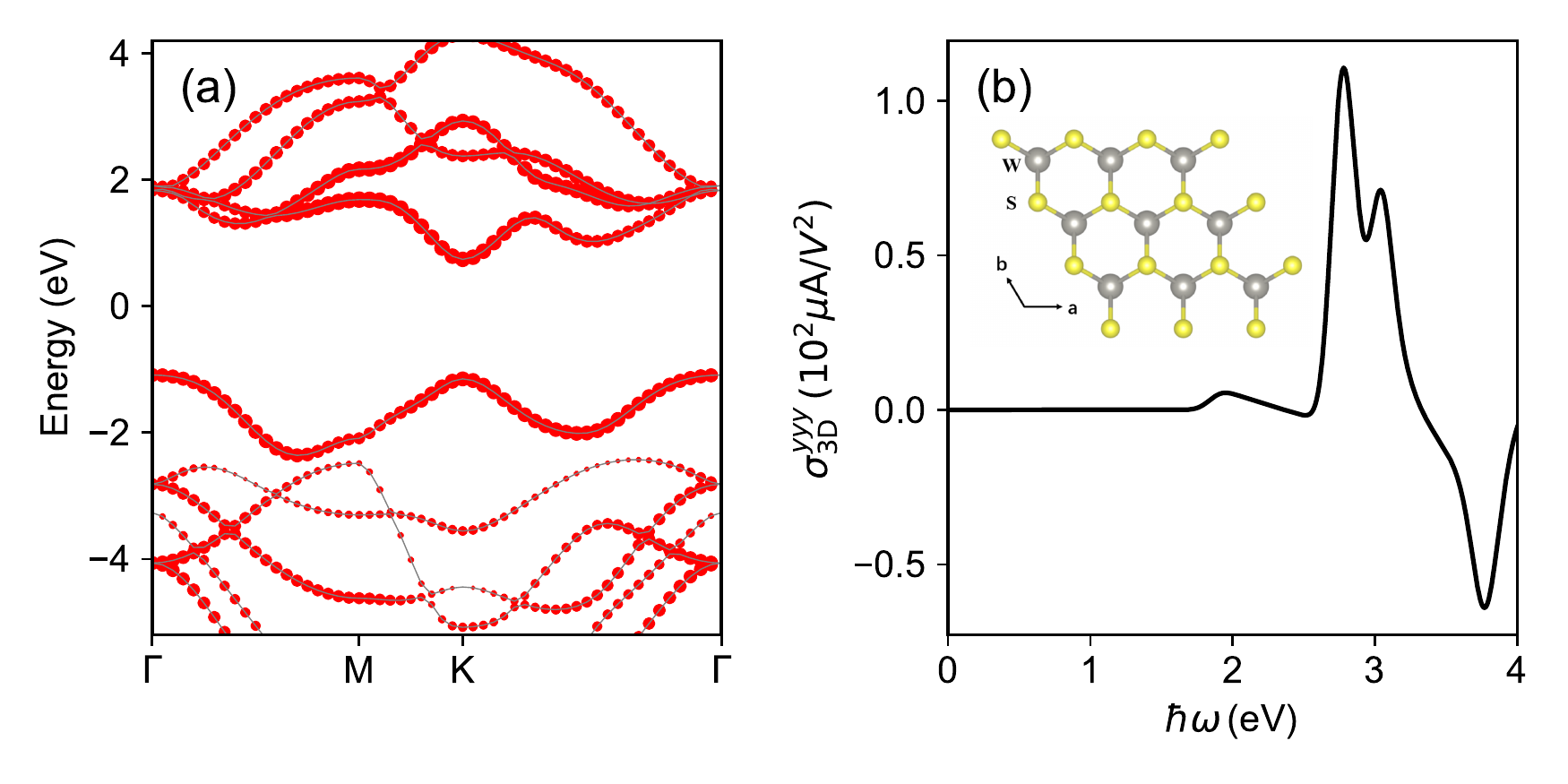}
	\caption{(a) The fat band structure of WS$_2$ for the W-5$d$ orbital. The size of the red circles represents the weight of the projection. (b) The $yyy$ component of the shift current in WS$_2$. The inset shows the crystal structure of WS$_2$.}
	\label{fig:WS2}
\end{figure}

We demonstrate the capabilities of PYATB's fat band and shift current functions using the WS$_2$ monolayer as an example. WS$_2$ is a transition metal dichalcogenide  that exhibits a range of intriguing optical properties, such as room temperature photoluminescence and optical Stark effect~\cite{gutierrez_extraordinary_2013, sie_valley-selective_2015}. In Fig.\ref{fig:WS2}(a), we show the fat band structure of WS$_2$, highlighting the contribution from the W-$5d$ orbital. The red circle sizes in the figure indicate the weight of the $5d$ orbitals.
Figure~\ref{fig:WS2}(b) illustrates the shift current conductivity of monolayer WS$2$. Since the system has $D_{6h}$ point group symmetry, it has only one independent component $\sigma^{yyy}$. These results are in good agreement with those obtained in Ref.~\cite{wang_first-principles_2019}.

The {\tt Input} file for the fat band and shift current calculations is given in the \ref{input_ws2}.

\begin{figure*}[htbp]
	\centering
	\includegraphics[width=0.8\textwidth]{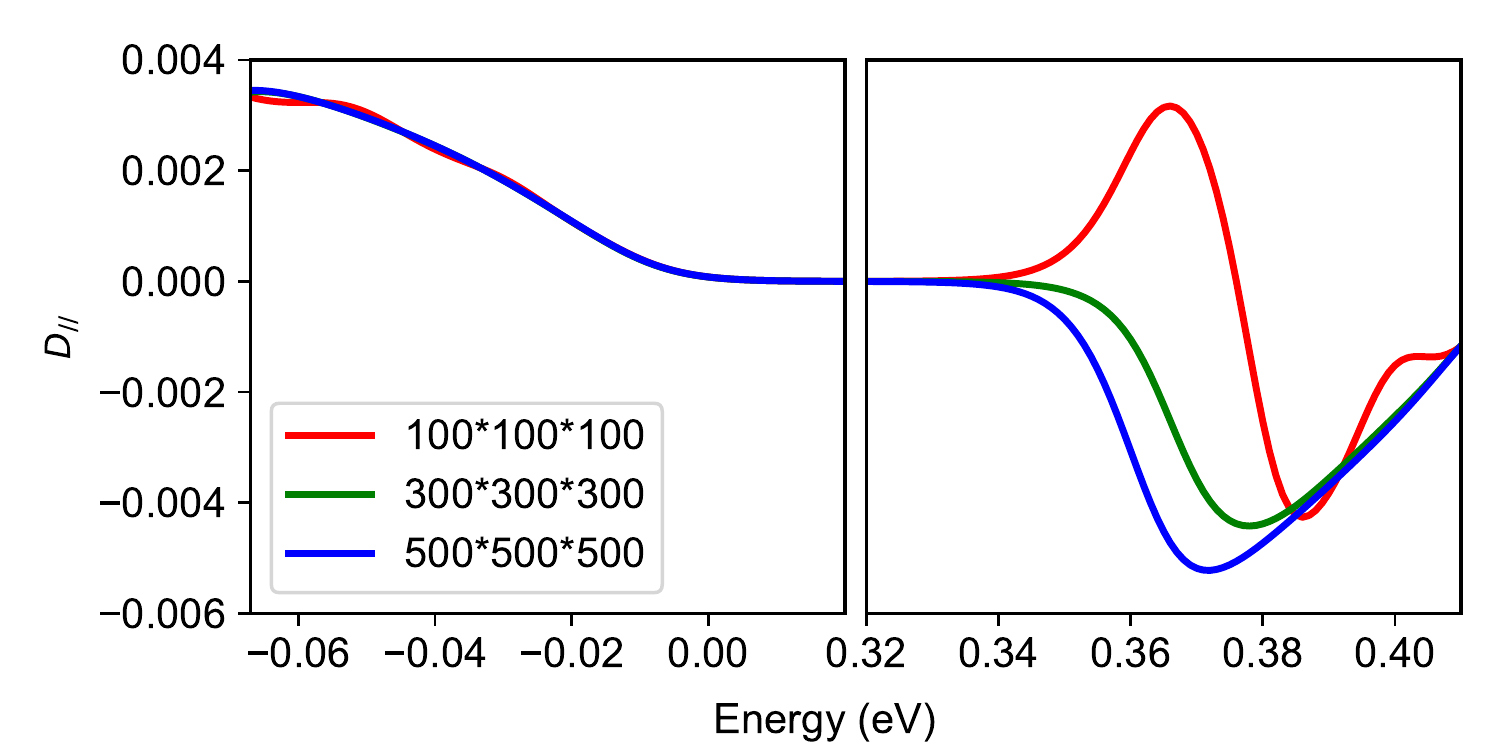}
	\caption{The Berry curvature dipole $D_{//}(E)$($=\frac{1}{2} D_{zz}(E)$) versus energy $E$ measured from the VBM using different
$\mathbf{k}$ point meshes.}
	\label{fig:Te}
\end{figure*}

\subsection{Berry curvature dipole}

In this example, we showcase the calculation of the Berry curvature dipole of trigonal Te using PYATB. In systems with time-reversal symmetry, a nonlinear anomalous Hall current exists due to a dipole moment induced by the unbalanced distribution of Berry curvature in $\mathbf{k}$-space caused by the breaking of inversion symmetry~\cite{sipe_second-order_2000}. The Berry curvature dipole  has recently gained attention due to its intriguing topological nature and its potential for photo-electric detections~\cite{du_quantum_2021, ma_observation_2019}.

To begin, we performed ab-initio calculations using the HSE functional implemented  in ABACUS
to generate the {\tt HR}, {\tt SR}, and {\tt rR} files.
We then used PYATB to calculate the Berry curvature dipole of trigonal Te by integrating the entire BZ with meshes of 100$\times$100$\times$100, 300$\times$300$\times$300, and 500$\times$500$\times$500 $\mathbf{k}$ points, respectively. When the Berry curvature at a mesh point exceeds the threshold, we refine the mesh to a size of 20$\times$20$\times$20. The resulting Berry curvature dipoles are shown in Fig.\ref{fig:Te}. Our results suggest that an extremely dense $\mathbf{k}$ points mesh is required to achieve convergence. Specifically, the results obtained with a 300-mesh size are in agreement with those previously reported in the literature using the same $\mathbf{k}$ points mesh~\cite{tsirkin_gyrotropic_2018}.

The relevant parameters for calculating the Berry curvature dipole in the {\tt Input} file are provided in \ref{input_te}.

\section{Summary}
\label{sec:summary}

The PYATB package is a user-friendly software that allows for the calculation of a broad range of physical properties of materials, such as band structures, associated topological properties, and optical properties. Its most notable advantage is that the ab initio tight-binding Hamiltonian can be naturally generated after the self-consistent calculations using NAO-based first-principles softwares, such as ABACUS, without the need to construct MLWF. This feature simplifies the computational process and ensures the correct symmetry for the systems. We demonstrate the capabilities of PYATB through a few illustrative examples. We hope that PYATB will become a convenient and efficient toolkit for studying the electronic structural properties of materials.

\section{Acknowledgments}

This work was supported by the Chinese National Science Foundation Grant Number 12134012, and the
Innovation Program for Quantum Science and Technology Grant Number 2021ZD0301200.
The numerical calculations were performed on the USTC HPC facilities.

\appendix

\section{Example input files}

\subsection{{\tt Input} file for NV center}
\label{input_nv}

\begin{verbatim}
  BANDUNFOLDING  # Specify the purpose of the calculation
  {
      # The STRU file includes the names of the NAO files.
      stru_file     STRU   # The file name for the crystal structure.
      ecut          400    # (eV), the cutoff energy for the plane wave basis of the PC.
      band_range    10 250 # Unfold the 10-th to 250-th bands of the supercell.
      m_matrix      -2 2 2, 2 -2 2, 2 2 -2 # m_ij, i,j = 1, 2, 3
      kpoint_mode   line
      kpoint_num    5    # There are 5 high symmetry k points in the line
      high_symmetry_kpoint
      0.500000  0.000000  0.500000 300  # X, kx, ky, kz, number of k points between X and W
      0.500000  0.250000  0.750000 300  # W
      0.500000  0.500000  0.500000 300  # L
      0.000000  0.000000  0.000000 300  # Gamma
      0.500000  0.000000  0.500000 1    # X
  }
\end{verbatim}

\subsection{{\tt Input} file for Bi$_2$Se$_3$}
\label{input_bi2se3}

\begin{verbatim}
  SPIN_TEXTURE
  {
      nband              78       # Specify the calculated energy band index.
      kpoint_mode        direct   # The k points are given in direct coordinates.
      kpoint_num         140
      kpoint_direct_coor
      0.010000  0.000000 0.000000 # Direct coordinates of the k point
      0.011187  0.003516 0.000000
      0.011279  0.006687 0.000000
      ...
      0.027084 -0.005339 0.000000
      0.028631 -0.002678 0.000000
  }

  WILSON_LOOP
  {
      occ_band      78  # Number of occupied energy bands.
      # To determine a plane of k-space requires an origin (k_start) and
      # two vectors that are not parallel to each other (k_vect1, k_vect2).
      k_start       0.0  0.0  0.5
      k_vect1       1.0  0.0  0.0
      k_vect2       0.0  0.5  0.0
      nk1           101 # number of points of the uniform divide k_vect1.
      nk2           101 # number of points of the uniform divide k_vect2.
  }
\end{verbatim}

\subsection{{\tt Input} file for MnSb$_2$Te$_4$}
\label{input_mnsb2te4}

\begin{verbatim}
  BAND_STRUCTURE
  {
      kpoint_mode                  line
      kpoint_num                   5
      high_symmetry_kpoint
      # Four numbers, the first three are special k-point coordinates
      # and the fourth is the number of k-points between this
      # special k-point and the next.
      0    0   0    200  # Gamma
      0    0   0.5  200  # Z
      0.5  0   0.5  200  # F
      0    0   0    200  # Gamma
      0.5  0   0    1    # L
  }

  FIND_NODES
  {
      # (eV), search for degenerate k-points with energies
      # in the 9.870 to 10.070 eV range.
      energy_range       9.870 10.070

      # Set the search space of k points.
      # Selecting a parallel hexahedron in k-space requires an
      # origin (k_start) and three vectors (k_vect1, k_vect2, k_vect3)
      # that are not parallel to each other. In this example, k_vect2
      # and k_vect3 are zero vectors, so the chosen search space is
      # k-line from (0.0, 0.0, -0.2) to (0.0, 0.0, 0.4).
      k_start            0.0 0.0 -0.2
      k_vect1            0.0 0.0  0.0
      k_vect2            0.0 0.0  0.0
      k_vect3            0.0 0.0  0.4

      # To start, insert the initial_grid into the search space.
      # Then, check each k-point for its band gap. If the band gap is less than
      # the initial_threshold, refine the k-point using the adaptive_grid located nearby.
      # After refinement, check the band gap of the refined k-point.
      # If it is less than the adaptive_threshold, output the k-point as a result.
      initial_grid       1  1  100
      initial_threshold  0.01  # (eV)
      adaptive_grid      1  1  20
      adaptive_threshold 0.001 # (eV)
  }

  CHIRALITY
  {
      k_vect        0.0000 0.0000 -0.0538 # k coordinates, determine its chirality.
      # unit is 1.0 / angstrom. Draw a spherical surface with the k-point as the
      # center and a radius of 0.02.
      radius        0.02
      point_num     100  # The number of k-points uniformly distributed on the sphere.
  }

  BERRY_CURVATURE
  {
      kpoint_mode                  mp
      # Selecting a parallel hexahedron in k-space requires an
      # origin (k_start) and three vectors (k_vect1, k_vect2, k_vect3)
      # that are not parallel to each other.
      k_start                      0 0 0
      k_vect1                      1 0 0
      k_vect2                      0 1 0
      k_vect3                      0 0 0.5
      # Number of grid points for uniformly dividing 3D k-Space.
      mp_grid                      300 300 50
  }

  CHERN_NUMBER
  {
      occ_band                     109 # Number of occupied energy bands.
      integrate_mode               Grid
      integrate_grid               100 100 1
      # When the Berry curvature of a k point is greater than the
      # threshold (adaptive_grid_threshold), increase the density of k-points
      # around the k point.
      adaptive_grid                20  20  1
      adaptive_grid_threshold      100
      # To determine a plane of k-space requires an origin (k_start) and
      # two vectors that are not parallel to each other (k_vect1, k_vect2).
      k_start                      0 0 0
      k_vect1                      1 0 0
      k_vect2                      0 1 0
  }
\end{verbatim}

\subsection{{\tt Input} file for CsPbI$_3$}
\label{input_cspbi3}

\begin{verbatim}
  JDOS
  {
      occ_band      37       # Number of occupied energy bands.
      omega         0.5  10  # (eV), hbar omega.
      domega        0.01     # energy interval.
      eta           0.2      # Gauss smearing parameters.
      grid          30 30 30 # k-space grid points.
  }

  OPTICAL_CONDUCTIVITY #  Calculate the optical conductivity as well as the dielectric functions
  {
      occ_band      37       # Number of occupied energy bands.
      omega         0.5  10  # (eV), hbar omega.
      domega        0.01     # energy interval.
      eta           0.2      # Gauss smearing parameters.
      grid          30 30 30 # k-space grid points.
  }
\end{verbatim}

\subsection{{\tt Input} file for WS$_2$}
\label{input_ws2}

\begin{verbatim}
  FAT_BAND
  {
      band_range      10 30
      stru_file       STRU    # The file name containing the crystal structure.
      kpoint_mode     line
      kpoint_num      4
      high_symmetry_kpoint
      # Four numbers, the first three are special k-point coordinates
      # and the fourth is the number of k-points between this
      # special k-point and the next.
      0.0000000000   0.0000000000   0.0000000000   20 # GAMMA
      0.5000000000   0.0000000000   0.0000000000   10 # M
      0.3333333333   0.3333333333   0.0000000000   25 # K
      0.0000000000   0.0000000000   0.0000000000   1  # GAMMA
  }

  SHIFT_CURRENT
  {
      occ_band         13     # Number of occupied energy bands.
      omega            0   4  # (eV), hbar omega.
      domega           0.01   # energy interval.
      smearing_method  1 # Gaussian smearing
      eta              0.1    # Gauss smearing parameter
      grid             1000 1000 1 # k-space grid points
  }
\end{verbatim}

\subsection{{\tt Input} file for Te}
\label{input_te}

\begin{verbatim}
  BERRY_CURVATURE_DIPOLE
  {
      omega                      9.474 10.074 # (eV) energy range.
      domega                     0.001        # energy interval.
      integrate_mode             Grid
      integrate_grid             500 500 500
      # When the Berry curvature of a k point is greater than the
      # threshold (adaptive_grid_threshold), increase the density of k-points
      # around the k point.
      adaptive_grid              20 20 20
      adaptive_grid_threshold    20000
  }
\end{verbatim}


\end{document}